%% file: SLIM_arxiv.tex
\documentclass[letterpaper]{article}
\usepackage[preprint]{aaai2027}

\makeatletter
\gdef\showauthors@on{T}
\makeatother
\nocopyright
\usepackage[hyphens]{url}
\usepackage{graphicx}
\usepackage{verbatim}
\usepackage{natbib}
\usepackage{amsmath,amssymb,bm}
\usepackage{booktabs}
\usepackage{xspace}
\usepackage{algorithm}
\usepackage{algpseudocode}

\newcommand{\name}{\textsc{SLIM}\xspace}

\title{\name: Learning an Operator for Material-Efficient 3D Lattice Refinement}
\author{
Yueze Zhu, Yu Xing, Tianyang Xue, Peng-Shuai Wang,\\
Jikai Liu, Haisen Zhao, Lin Lu\corresponding
}
\affiliations{}

\begin{document}

\maketitle

\begin{abstract}
\input{src/abstract}
\end{abstract}

\input{src/1-introduction}
\input{src/2-relatedwork}

\input{src/3-background}
\input{src/4-method}
\input{src/5-result}
\input{src/6-discussion}
\newpage

\input{src/appendix}
\bibliography{src/ref}
\end{document}

%% file: src/abstract.tex
Lattice inverse-design methods can generate cells that match prescribed effective properties, but property feasibility does not imply material efficiency. Removing redundant material from an existing lattice while preserving its homogenized elasticity ordinarily requires solving a separate constrained topology optimization problem for each cell, making large-scale refinement computationally expensive. We introduce SLIM (Structural Lightweighting via Implicit Mapping), a neural lattice-refinement model that replaces this per-instance iterative optimization with a single forward prediction. SLIM uses a dual-octree implicit network to predict the signed-distance field of a lightweight lattice and is trained on input--output pairs generated by property-constrained topology optimization. To improve stiffness preservation, we use homogenization sensitivities computed on the optimized target to estimate how residual voxel errors in a prediction affect its homogenized stiffness, and penalize this estimated change during training. Once trained, SLIM requires no topology optimization, homogenization analysis, or sensitivity evaluation during prediction. On the held-out test set of 906 lattices, SLIM achieves a mean relative volume-fraction reduction of \(15.90\%\), a lightweighting success rate of \(90.51\%\), and a mean normalized aggregate TO-target error of \(0.636\%\), while reducing the average per-lattice generation time from \(2516.0\) s (\(41.9\) min) with direct TO to approximately \(0.5\) s. These results show that SLIM provides a fast neural surrogate for repeated topology-optimization-based refinement of 3D lattices.

%% file: src/1-introduction.tex
\section{Introduction}
Lattice structures and architected cellular materials provide unusual
stiffness--density trade-offs through geometric design and are widely used
in lightweight engineering~\cite{fleck2010microarchitectured,
schaedler2016architected,kadic2019metamaterials}. In multi-scale design, heterogeneous lattice cells are assigned to different regions of a structure to realize spatially varying homogenized elastic properties \cite{panetta_elastic_2015, schumacher_microstructures_2015, zhu_two-scale_2017, wu_topology_2021, white2019multiscale, groen2018homogenization}. Recent learning-based inverse-design methods can efficiently generate lattice cells that match prescribed elasticity tensors. Recent learning-based inverse-design methods can efficiently generate
lattice cells that match prescribed mechanical responses
~\cite{kumar2020spinodoid,kollmann2020deeplearning,ha2023rapid,bastek_inverting_2022,zheng_unifying_2023,xue_mind_2025,yang_guided_2026}. However, property feasibility does not imply material efficiency. Because the inverse structure--property relationship is one-to-many, geometrically distinct lattices can exhibit similar effective properties while using different amounts of material. Consequently, a generated or selected lattice may satisfy its prescribed elastic response while still containing removable material.

We therefore consider post-generation lattice refinement: given an existing property-feasible lattice, remove redundant material while retaining its input-specific homogenized elastic response. This task differs from conventional inverse design, which synthesizes a new geometry from target properties but does not refine a supplied geometry relative to its existing response. Property-constrained topology optimization can solve the refinement problem explicitly, but it requires repeated homogenization, sensitivity evaluation, and iterative geometry updates for every input lattice \cite{sigmund_materials_1994,andreassen_how_2014,wu_topology_2021}. The resulting per-instance cost becomes prohibitive when many distinct cells must be refined \cite{zhang_optimized_2023,xing2026gmt}. Faster alternatives such as uniform thinning or morphological erosion are not property-aware and may remove mechanically important material.

A constrained topology-optimization solver maps each property-feasible input lattice $V^{\mathrm{in}}$ to a lower-volume solution $V^{\mathrm{TO}}$ that satisfies the prescribed elasticity constraint. Computing this solution requires repeated homogenization, sensitivity analysis, and geometry updates for every input. We instead train a neural network $\mathcal{R}_{\theta}$ to approximate this input-to-refinement mapping,
\[
\mathcal{R}_{\theta}(V^{\mathrm{in}}) \approx V^{\mathrm{TO}}.
\]
Once trained, the network predicts a refined lattice in a single forward pass, avoiding iterative topology optimization for each new input.

We introduce SLIM, Structural Lightweighting via Implicit Mapping, a neural lattice-refinement model trained on pairs of property-feasible lattices and their topology-optimized lightweight counterparts. Given an input lattice, SLIM uses a dual-octree implicit network to predict the continuous signed-distance field of its refinement target. A geometric reconstruction loss alone measures how closely the predicted geometry matches the target, but does not account for how individual voxel errors affect mechanical properties. We therefore compute homogenization sensitivities on each optimized target and use them to estimate how the remaining voxel differences between the prediction and the target change the homogenized stiffness. This estimated stiffness deviation is penalized during training. Because the sensitivities depend on the current structure, they are evaluated on the optimized target, where the prediction is expected to lie.

\input{figures/pipeline}

Our main contributions are:
\begin{itemize}
    \item We formulate post-generation lattice lightweighting as an
    input-conditioned geometry-refinement problem, distinguishing the
    refinement of an existing property-feasible lattice from conventional
    property-conditioned lattice generation.

    \item We introduce SLIM, an input-conditioned dual-octree implicit
    model that learns the geometry transformations represented by
    boundary- and property-constrained topology-optimization targets.
    We further propose a target-state homogenization-sensitivity loss
    that uses sensitivities evaluated at the optimized target to penalize
    the estimated stiffness deviations caused by residual voxel errors.

    \item On 906 held-out lattices, SLIM reduces the volume
fraction by $15.90\%$ on average, with a $90.51\%$ success
rate and a mean normalized TO-target error of $0.636\%$,
while cutting the per-lattice generation time from $2516.0$
s to $0.5$ s, a $5032\times$ speedup.
\end{itemize}

%% file: figures/pipeline.tex
\begin{figure}[t]
    \centering
    \includegraphics[width=\columnwidth]{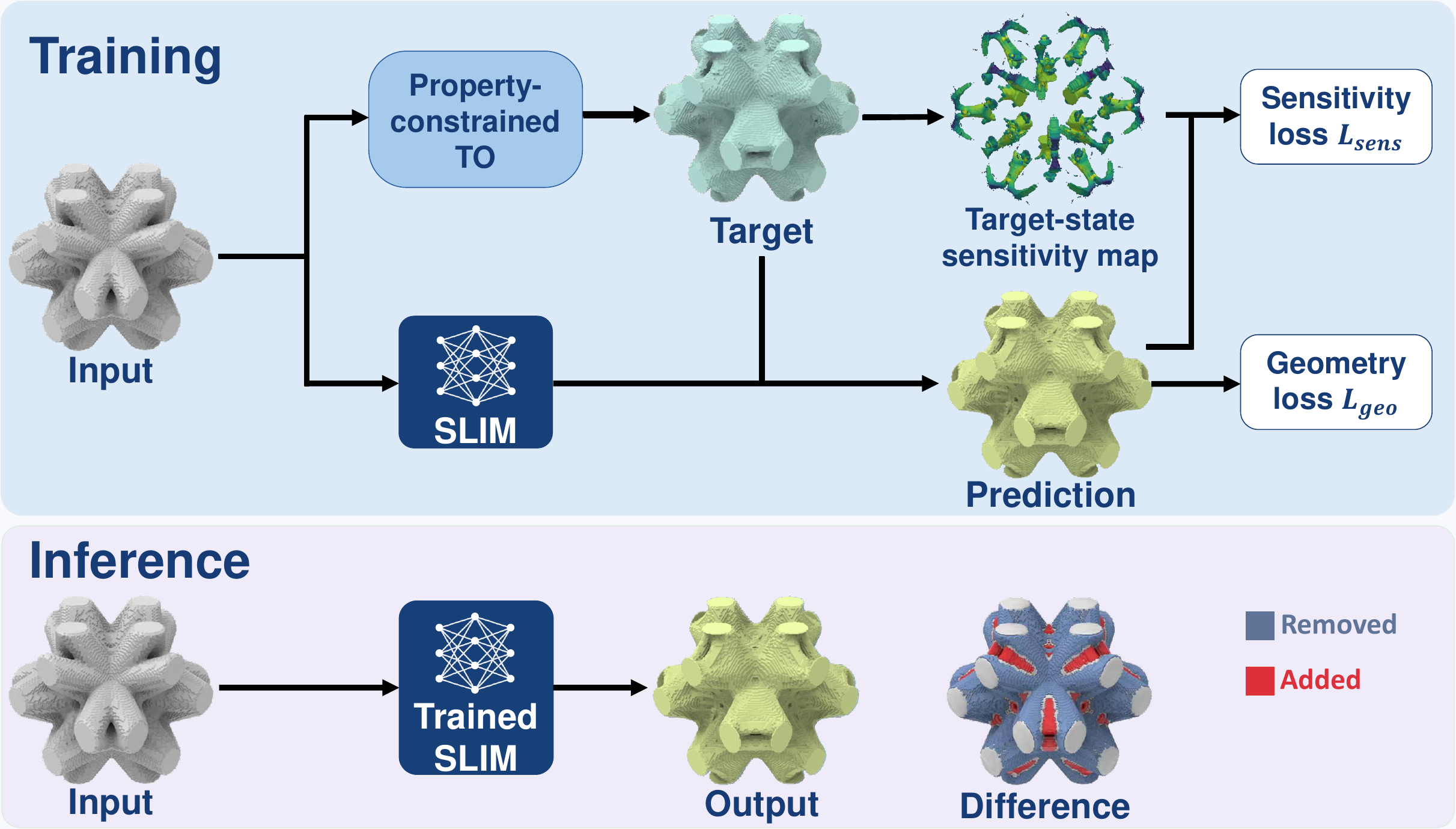}
    \caption{\textbf{Overview of SLIM.}
During offline training, property-constrained topology optimization
(TO) produces a lightweight target for each input lattice, and
homogenization sensitivities evaluated at the target provide the
target-state sensitivity map. SLIM is trained to reconstruct the
TO target using the geometric loss $\mathcal{L}_{\mathrm{geo}}$
and the sensitivity loss $\mathcal{L}_{\mathrm{sens}}$.
At inference, the trained model predicts a refined lattice from
the input in a single forward pass, without topology optimization,
homogenization, or sensitivity evaluation.
In the difference view, blue and red denote material removed from
and added to the input, respectively.}
    \label{fig:slim_overview}
\end{figure}

%% file: src/2-relatedwork.tex
\section{Related Work}

\subsection{Property-Driven Microstructure Design and Refinement}
\label{sec:related_property_refinement}

Multiscale lattice design assigns microstructures to different regions of a macroscopic structure to realize spatially varying material behavior \cite{wu_topology_2021,lee_datadriven_2024,xing2026gmt}. In conventional top-down workflows, target effective properties are specified at the macroscale, and compatible unit cells are selected or generated accordingly \cite{panetta_elastic_2015,schumacher_microstructures_2015,zhu_two-scale_2017}. Classical inverse homogenization iteratively optimizes unit-cell geometry or density to match prescribed elastic properties, optionally under volume constraints \cite{sigmund_materials_1994,andreassen_how_2014,wu_topology_2021}. More broadly, homogenization-based and level-set formulations established
the main numerical paradigms for changing structural topology under
mechanical objectives and constraints~\cite{bendsoe1988generating,
wang2003levelset,allaire2004structural}.
Data-driven inverse-design methods accelerate this process by learning mappings from target properties to feasible geometries, using regression, latent generative models, or diffusion models \cite{bastek_inverting_2022,zheng_unifying_2023,yang_guided_2026,xue_mind_2025}.

These methods predominantly formulate microstructure design as property-conditioned synthesis.  
Because the inverse structure--property map is generally one-to-many, matching a target response does not identify a unique geometry and does not by itself favor material-efficient solutions~\cite{regenwetter2022generative, ha2023rapid}.
SLIM instead addresses post-generation refinement: it takes an existing property-feasible lattice as input and learns from topology-optimization pairs to predict a lower-volume counterpart relative to its input-specific elastic response. The prescribed boundary band is retained explicitly, whereas elastic-property preservation is evaluated after prediction.

\subsection{Learning-Based Surrogates for Structural Optimization}
\label{sec:related_surrogates}

Conventional numerical topology optimization solves each design instance through repeated homogenization, sensitivity evaluation, and geometry updates, resulting in substantial computational cost for high-resolution three-dimensional structures \cite{xing2024shell,andreassen_how_2014,wu_topology_2021}. Learning-based topology-optimization methods seek to avoid this per-instance cost by predicting optimized layouts from loads, supports, volume constraints, or physical fields \cite{yu_deep_2019,nie_topologygan_2021,maze_diffusion_2023}. Early learning-based approaches formulated topology prediction as image segmentation or one-shot generation, while later reviews emphasized the need to assess generalization and physical optimality rather than inference speed alone~\cite{sosnovik2019neural,cang2019oneshot,woldseth2022ann}. Neural parameterizations can also represent continuous density fields and improve geometric compatibility, but their parameters are typically optimized separately for each design instance \cite{chandrasekhar_tounn_2021,chen_multiscale_2025,zehnder2021ntopo}. More recent methods combine direct prediction with implicit neural fields to support iteration-free or resolution-flexible topology synthesis \cite{hu_iftonir_2024,nobari_3d_2024}. These approaches primarily map problem specifications to newly synthesized structures rather than learn a shared transformation from a complete feasible geometry to its refined counterpart.
Neural implicit representations encode geometry as the level set of a continuous function and can represent complex topology without an explicit output mesh~\cite{mescheder2019occupancy,chen2019implicit, chibane2020ifnets,park_deepsdf_2019}. For detailed three-dimensional structures, octree-based networks concentrate computation near geometrically relevant regions. Dual-octree graph networks further use hierarchical features to reconstruct continuous volumetric fields efficiently \cite{wang_dualoctree_2022}. SLIM adopts this representation to model sparse, high-resolution lattice geometry; the implicit architecture is an enabling component rather than the primary contribution.

SLIM learns a shared geometry-to-geometry mapping from pairs generated by boundary- and property-constrained topology optimization. Unlike specification-to-geometry predictors, it conditions on the complete input lattice and approximates its input-specific refinement. A purely geometric reconstruction loss, however, does not distinguish errors by their influence on homogenized stiffness. SLIM therefore augments SDF supervision with signed homogenization sensitivities evaluated at the optimization targets, emphasizing occupancy errors according to their estimated mechanical consequences.

%% file: src/3-background.tex
\section{Problem Formulation}
\label{sec:problem}

Let $\mathbf V^{\mathrm{in}}\in\{0,1\}^{N\times N\times N}$ denote an
existing property-feasible lattice. We take its homogenized elasticity
tensor as the input-specific response to be preserved:
\begin{equation}
    \mathbf C^{\mathrm{tar}}
    :=
    \mathbf C^H(\mathbf V^{\mathrm{in}}).
\end{equation}
Our goal is to remove redundant material while retaining this effective
elastic response and the prescribed assembly boundarys.

Let $\mathcal B$ denote the boundary region that must remain unchanged. For
a binary lattice $\mathbf V$, we define its volume fraction and relative
elasticity-tensor error as
\begin{equation}
    \operatorname{VF}(\mathbf V)
    =
    \frac{1}{N^3}\sum_{i=1}^{N^3}V_i,
    \quad
    e_C(\mathbf V,\mathbf V^{\mathrm{in}})
    =
    \frac{
        \|\mathbf C^H(\mathbf V)-\mathbf C^{\mathrm{tar}}\|_F
    }{
        \|\mathbf C^{\mathrm{tar}}\|_F
    }.
\end{equation}
The desired refinement is then formulated as
\begin{equation}
\begin{aligned}
\mathbf V^{\star}
\in
\underset{\mathbf V\in\{0,1\}^{N\times N\times N}}{\arg\min}
\quad&
\operatorname{VF}(\mathbf V)
\\
\mathrm{subject\ to}\quad&
e_C(\mathbf V,\mathbf V^{\mathrm{in}})
\leq \epsilon_C,
\\
&
\mathbf V|_{\mathcal B}
=
\mathbf V^{\mathrm{in}}|_{\mathcal B},
\end{aligned}
\label{eq:task_constraints}
\end{equation}
where $\epsilon_C$ is the prescribed relative elasticity-tensor tolerance.

Solving Eq.~\ref{eq:task_constraints} independently for every input is
computationally expensive. We therefore use the numerical topology-
optimization procedure $\mathcal T_{\mathrm{TO}}$ to generate a teacher
target
\begin{equation}
    \mathbf V^{\mathrm{TO}}
    =
    \mathcal T_{\mathrm{TO}}(\mathbf V^{\mathrm{in}}),
\end{equation}
and train a neural refinement model to approximate the resulting mapping:
\begin{equation}
\mathbf{V}^{\mathrm{pred}}
=
\mathcal{R}_{\theta}(\mathbf{V}^{\mathrm{in}})
\approx
\mathbf{V}^{\mathrm{TO}}.
\label{eq:learned_refinement}
\end{equation}

Here, $\mathcal{R}_{\theta}$ denotes the complete prediction
pipeline, including implicit-field prediction, voxelization,
and deterministic boundary restoration;
$\mathbf{V}^{\mathrm{pred}}$ is its final binary output.
Because Eq.~\ref{eq:task_constraints} may admit multiple valid geometries
and the numerical optimization is non-convex,
$\mathbf V^{\mathrm{TO}}$ is treated as a teacher solution rather than a
guaranteed global optimum.

The boundary constraint is satisfied by restoring the prescribed region
after voxelization. Material reduction and elasticity preservation are not
formally guaranteed by the learned model and are therefore measured on the
test set. In particular, $C^H(\mathbf{V}^{\mathrm{pred}})$ is computed by post-hoc numerical homogenization and is not used to modify the predicted geometry. The target-generation procedure and learned model are described in Section~\ref{sec:method}, with complete optimization details provided in Appendix~B.

%% file: src/4-method.tex
\section{Method}
\label{sec:method}

\paragraph{Overview.}
Given a property-feasible lattice $V^{\mathrm{in}}$, SLIM predicts a
lightweight geometry that approximates a topology-optimized target
$V^{\mathrm{TO}}$. The method has three stages. First, we generate paired
training data by refining each input lattice through property-constrained
topology optimization. Second, a dual-octree implicit network predicts the
signed-distance field (SDF) of the optimized target. Third, in addition to
geometric reconstruction, we use homogenization sensitivities evaluated at
the optimized target to estimate how the remaining voxel errors in a
prediction affect its homogenized stiffness. This estimate is used as a
training loss. The network does not perform topology optimization or
homogenization during prediction.

\subsection{Paired Refinement Targets}

For each input lattice, we generate a lightweight teacher target using
the input-specific tensor $\mathbf C^{\mathrm{tar}}$ defined in
Section~\ref{sec:problem}. The optimization searches for a lower-volume geometry that matches this tensor within the prescribed tolerance while retaining the fixed interface region.

Because the corresponding binary problem is large and non-convex, we use a
nested numerical procedure. An outer bisection search progressively lowers
the admissible volume fraction, while an inner MMA solve minimizes the
homogenized-tensor mismatch for each trial bound. Selected continuous
iterates are binarized and re-evaluated by numerical homogenization. Among
all binary checkpoints satisfying the volume and elasticity tolerances, we
select
\begin{equation}
    V^{\mathrm{TO}}
    =
    \arg\min_{V\in\mathcal Q_{\mathrm{feas}}}\mathrm{VF}(V).
\end{equation}

Inputs for which no strictly lower-volume feasible checkpoint found are
excluded. This procedure produces training pairs
$(V^{\mathrm{in}},V^{\mathrm{TO}})$. Because the inner optimization is
non-convex and binarization can change both stiffness and volume,
$V^{\mathrm{TO}}$ is treated as a numerical teacher solution rather than a
guaranteed global optimum. Appendix~B provides the complete optimization,
sensitivity, binarization, and bisection procedures.

\subsection{Implicit Lattice Refinement}

We convert the target lattice $V^{\mathrm{TO}}$ into an SDF
$d^{\mathrm{TO}}$, whose zero level set represents the target surface. The
input lattice is represented by oriented surface samples and encoded as an
adaptive octree $\mathcal O(V^{\mathrm{in}})$. A dual-octree graph network
predicts the target implicit field as
\begin{equation}
    \widehat d_{\theta}(\mathbf x;V^{\mathrm{in}})
    =
    F_{\theta}\!\left(\mathbf x,\mathcal O(V^{\mathrm{in}})\right)
    \approx
    d^{\mathrm{TO}}(\mathbf x),
\end{equation}
where $\mathbf x\in\mathbb R^3$ is a query location.

Following the multi-scale supervision of the dual-octree representation,
the geometric objective combines octree-split classification, SDF
regression, and SDF-gradient matching:
\begin{equation}
    \mathcal L_{\mathrm{geo}}
    =
    \mathcal L_{\mathrm{split}}
    +\lambda_{\mathrm{sdf}}\mathcal L_{\mathrm{sdf}}
    +\lambda_{\mathrm{grad}}\mathcal L_{\mathrm{grad}}.
\end{equation}
This objective supervises the optimized geometry, but it measures errors
according to geometric discrepancy and does not directly account for their
effects on homogenized stiffness.

\subsection{Target-State Homogenization-Sensitivity Loss}

To connect geometric prediction errors with mechanical behavior, we convert
the predicted SDF into a differentiable voxel occupancy. Let $M=N^3$ denote
the number of voxels. At voxel center $\mathbf x_i$, with negative SDF
values denoting solid material, we define
\begin{equation}
    \widehat\rho_{\theta,i}
    =
    \operatorname{sigmoid}\!\left(
        -\frac{\widehat d_{\theta}(\mathbf x_i)}{\tau}
    \right),
    \qquad
    \Delta\boldsymbol\rho
    =
    \widehat{\boldsymbol\rho}_{\theta}
    -\boldsymbol\rho^{\mathrm{TO}},
\end{equation}
where $\tau$ controls the sharpness of the SDF-to-occupancy conversion and
$\boldsymbol\rho^{\mathrm{TO}}$ is the binary occupancy of the optimized
target.

Let $\mathbf c^H(\boldsymbol\rho)\in\mathbb R^K$ collect the $K$
homogenized-stiffness components used in training; their Voigt ordering and
symmetry treatment are specified in the appendix. For a structural state
$s\in\{\mathrm{in},\mathrm{TO}\}$, we define
\begin{equation}
    \mathbf J^s
    =
    \left.
    \frac{\partial\mathbf c^H}{\partial\boldsymbol\rho}
    \right|_{\boldsymbol\rho^s}
    \in\mathbb R^{K\times M},
    \qquad
    J^s_{k,i}
    =
    \left.
    \frac{\partial c^H_k}{\partial\rho_i}
    \right|_{\boldsymbol\rho^s}.
\end{equation}
Here, $J^s_{k,i}$ is the marginal change in stiffness component $k$ caused
by a small perturbation of voxel $i$, with the equilibrium response
evaluated at the complete structural state $\boldsymbol\rho^s$. It is not
the amount of stiffness already contributed by that voxel. Although
$\boldsymbol\rho^{\mathrm{in}}$ and $\boldsymbol\rho^{\mathrm{TO}}$ are
defined on the same voxel grid, material removal changes the equilibrium
fields and load paths; therefore,
$\mathbf J^{\mathrm{in}}\neq\mathbf J^{\mathrm{TO}}$ in general.

Because the network is trained to reconstruct the optimized target, we
evaluate the Jacobian at $\boldsymbol\rho^{\mathrm{TO}}$ and linearize the
homogenized response around this state:
\begin{equation}
    \mathbf c^H(\widehat{\boldsymbol\rho}_{\theta})
    -
    \mathbf c^H(\boldsymbol\rho^{\mathrm{TO}})
    \approx
    \mathbf J^{\mathrm{TO}}
    \left(
        \widehat{\boldsymbol\rho}_{\theta}
        -
        \boldsymbol\rho^{\mathrm{TO}}
    \right)
    =
    \mathbf J^{\mathrm{TO}}\Delta\boldsymbol\rho.
\end{equation}
This approximation estimates how the voxel errors remaining near the
training target change its homogenized stiffness. We therefore define
\begin{equation}
    \mathcal L_{\mathrm{sens}}
    =
    \left\|
        \mathbf J^{\mathrm{TO}}\Delta\boldsymbol\rho
    \right\|_2^2.
\end{equation}
For stiffness component $k$, the estimated residual is
\begin{equation}
    \left[
        \mathbf J^{\mathrm{TO}}\Delta\boldsymbol\rho
    \right]_k
    =
    \sum_{i=1}^{M}
    J^{\mathrm{TO}}_{k,i}\Delta\rho_i.
\end{equation}
The signed sum retains reinforcing and canceling effects among voxel errors,
unlike assigning each voxel an independent nonnegative importance weight.

The optimized target already satisfies
$e_C(\boldsymbol\rho^{\mathrm{TO}},
\boldsymbol\rho^{\mathrm{in}})\leq\epsilon_C$. Reducing the
prediction-to-target residual therefore helps preserve the input response,
up to the residual mismatch of the numerical teacher. Because the Taylor
approximation is local, $\mathcal L_{\mathrm{sens}}$ encourages rather than
guarantees satisfaction of the original elasticity constraint.

For efficiency, we estimate the matrix-vector product in the sensitivity
loss using $32768$ sampled voxel centers per training example. The
sampling strategy, scaling, stiffness-component ordering, and Jacobian
construction are detailed in the appendix.

The complete training objective is
\begin{equation}
    \mathcal L
    =
    \mathcal L_{\mathrm{geo}}
    +
    \lambda_J\mathcal L_{\mathrm{sens}}.
\end{equation}
The target Jacobian is precomputed offline and held fixed during network
training. Gradients pass through the soft occupancy to the predicted SDF;
the network neither predicts the Jacobian nor performs homogenization during
training.

%% file: src/5-result.tex
\section{Experiments}
\label{sec:experiments}

We evaluate SLIM in terms of its fidelity to direct topology
optimization (TO), inference efficiency and sensitivity-loss contribution. All physical properties are recomputed from the final
binary predictions by numerical homogenization.

\subsection{Dataset and Metrics}
\label{sec:dataset_metrics}

Our dataset consists of library lattices paired with
boundary- and property-constrained TO targets. We retain inputs for which
TO finds a feasible target with a lower volume fraction and split the
resulting pairs by input-lattice identity. The held-out test set contains
906 lattices. Complete dataset statistics and construction details are
provided in the appendix.

For test sample $n$, the relative volume-fraction reduction is
\begin{equation}
r_{\mathrm{VF},n}
=
\frac{
\mathrm{VF}_{n}^{\mathrm{in}}
-
\mathrm{VF}_{n}^{\mathrm{pred}}
}{
\mathrm{VF}_{n}^{\mathrm{in}}
}.
\label{eq:slim_objective}
\end{equation}
We report its test-set mean as $r_{\mathrm{VF}}$. Effective Young's
modulus $E$ is additionally reported as
interpretable property summaries.

Following~\cite{yang_guided_2026,xue_mind_2025}, we measure agreement with
the paired TO target using
\begin{equation}
e_{ij,n}^{\mathrm{TO}}
=
\frac{
\left|
C_{ij,n}^{\mathrm{pred}}
-
C_{ij,n}^{\mathrm{TO}}
\right|
}{
C_{ij}^{\max}-C_{ij}^{\min}
},
\qquad ij\in\{11,12,44\},
\label{eq:yang_component_error}
\end{equation}
where the normalization ranges are computed from the training targets.
The aggregate error is
\begin{equation}
e_{\mathrm{all},n}^{\mathrm{TO}}
=
\frac{
e_{11,n}^{\mathrm{TO}}
+
e_{12,n}^{\mathrm{TO}}
+
e_{44,n}^{\mathrm{TO}}
}{3}.
\label{eq:yang_all_error}
\end{equation}

Finally, we define the lightweighting success rate as
\begin{equation}
r_{\mathrm{succ}}
=
\frac{1}{N}
\sum_{n=1}^{N}
\mathbb{I}
\left[
e_{\mathrm{all},n}^{\mathrm{TO}}<1.25\%
\ \land\
\mathrm{VF}_{n}^{\mathrm{pred}}
<
\mathrm{VF}_{n}^{\mathrm{in}}
\right].
\label{eq:success_rate}
\end{equation}
A prediction is successful if it removes material while remaining within
the prescribed normalized error of its TO target.

\input{tables/lightweighting}

\subsection{Lightweighting Performance and Structure-Dependent
Refinement}
\label{sec:lightweighting_performance}

\input{figures/testset}

\input{figures/gallary}

\input{figures/assembly}

We first assess whether SLIM reproduces the lightweighting behavior represented by property-constrained topology optimization. Direct TO serves as the numerical teacher rather than as a competing learned predictor. Table~\ref{tab:lightweighting} reports the main test-set results together with the $\mathcal{L}_{\mathrm{sens}}$ ablation, which is analyzed separately in Section~\ref{sec:sensitivity_ablation}.
Because a larger volume reduction can also result from violating the target-property constraint, we treat the joint success rate $r_{\mathrm{succ}}$, rather than volume reduction alone, as the primary performance measure.

On the 906 held-out lattices, SLIM achieves a mean relative volume-fraction reduction of 15.90\%, a success rate of 90.51\%, and a mean normalized aggregate TO-target error $e_{\mathrm{all}}^{\mathrm{TO}}$ of 0.636\%.
The Direct-TO targets reduce the volume fraction by 14.90\% on average and attain a success rate of 100\% by construction. The slightly larger mean reduction produced by SLIM should not be interpreted as outperforming Direct TO: unlike the teacher optimization, the learned prediction does not explicitly enforce the elasticity constraint and can occasionally remove excess material. Rather, the 90.51\% success rate indicates that SLIM reproduces the teacher's refinement behavior for most test inputs while achieving a net material reduction in a single forward pass.

The test-set mean, however, conceals substantial variation in how much material can be removed from different lattices. Figure~2 shows that, at comparable input volume fractions, lower-stiffness lattices tend to admit larger reductions, whereas higher-stiffness lattices leave less room for lightweighting. The two annotated inputs both have $\mathrm{VF}_{\mathrm{in}}=51.8\%$, but are refined to $\mathrm{VF}_{\mathrm{pred}}=51.0\%$ and $40.8\%$, respectively. This trend is consistent with a stiffness-utilization interpretation: lower effective stiffness suggests more removable material under the same refinement objective, while higher stiffness indicates less apparent redundancy. Thus, volume fraction alone is insufficient to determine the refinement, motivating the complete input geometry as the conditioning signal. This is an empirical association within the present test distribution rather than a universal stiffness-density law, and property fidelity should be interpreted jointly with the success rate in Table~\ref{tab:lightweighting}.

Figure~\ref{fig:lattice_gallery} provides a geometric view of this input-dependent behavior. The voxel-wise difference views show that the changes are spatially nonuniform: material is removed from extended low-utility regions, while localized additions appear where the predicted geometry differs from the TO target. Thus, for these examples, SLIM performs a geometry-dependent redistribution with a net volume decrease rather than applying uniform thinning or morphological erosion.

Finally, Figure~\ref{fig:assembly_results} evaluates whether the learned cell-wise mapping can be composed within a heterogeneous assembly. We apply SLIM independently to 392 lattice cells and restore the prescribed outermost-voxel layer of every cell, thereby maintaining matching interfaces by construction. The cell-wise maps show that the amount of material removed varies across the assembly, consistent with the structure-dependent behavior observed in Figure~\ref{fig:reduction_map}. The total volume fraction decreases from 53.5\% to 46.3\%, corresponding to a 13.4\% relativ reduction in material. This experiment demonstrates geometric compatibility and cell-wise applicability in a heterogeneous arrangement; it does not constitute a constraint-certified evaluation of the complete assembly-level mechanical response.

Taken together, the results support three conclusions. SLIM reproduces the Direct-TO refinement behavior for most held-out inputs, the achievable material reduction depends on the input geometry and stiffness rather than on volume fraction alone, and the learned refinement can be applied cell by cell while preserving prescribed interfaces.

\subsection{Runtime}
\label{sec:runtime}
\input{tables/time}

We compare direct TO and SLIM on identical hardware and group the test lattices by input volume fraction. For TO, the reported runtime covers the complete target-generation procedure, including all outer volume-fraction trials, the accumulated inner optimization iterations, and the associated homogenization and sensitivity computations. ``TO iter.'' denotes the total number of inner optimization iterations accumulated over these trials. For SLIM, the runtime includes point sampling, octree construction, network inference, SDF querying, and binary voxel reconstruction. Post-hoc numerical homogenization is excluded because it is used only to evaluate the final prediction and does not participate in generating or
modifying it.

As shown in Table~\ref{tab:runtime}, direct TO requires $2282.8$--$2802.3$ s per lattice across the five input-volume-fraction intervals, whereas SLIM consistently completes the entire prediction pipeline in approximately $0.5$ s. Over the full test set, direct TO requires an average of $30{,}459.2$ accumulated optimization iterations and $2516.0$ s ($41.9$ min) per lattice. SLIM therefore achieves an average speedup of $5032\times$, with the speedup remaining above $4500\times$ in every volume-fraction interval. These results show that the computational advantage of SLIM is consistent across the evaluated density range, reducing repeated lattice refinement from a per-instance iterative solve lasting tens of minutes to a sub-second neural prediction pipeline.

\subsection{Ablation of the Target-State Sensitivity Loss}
\label{sec:sensitivity_ablation}

\begin{figure}[t]
    \centering
    \includegraphics[width=\linewidth]{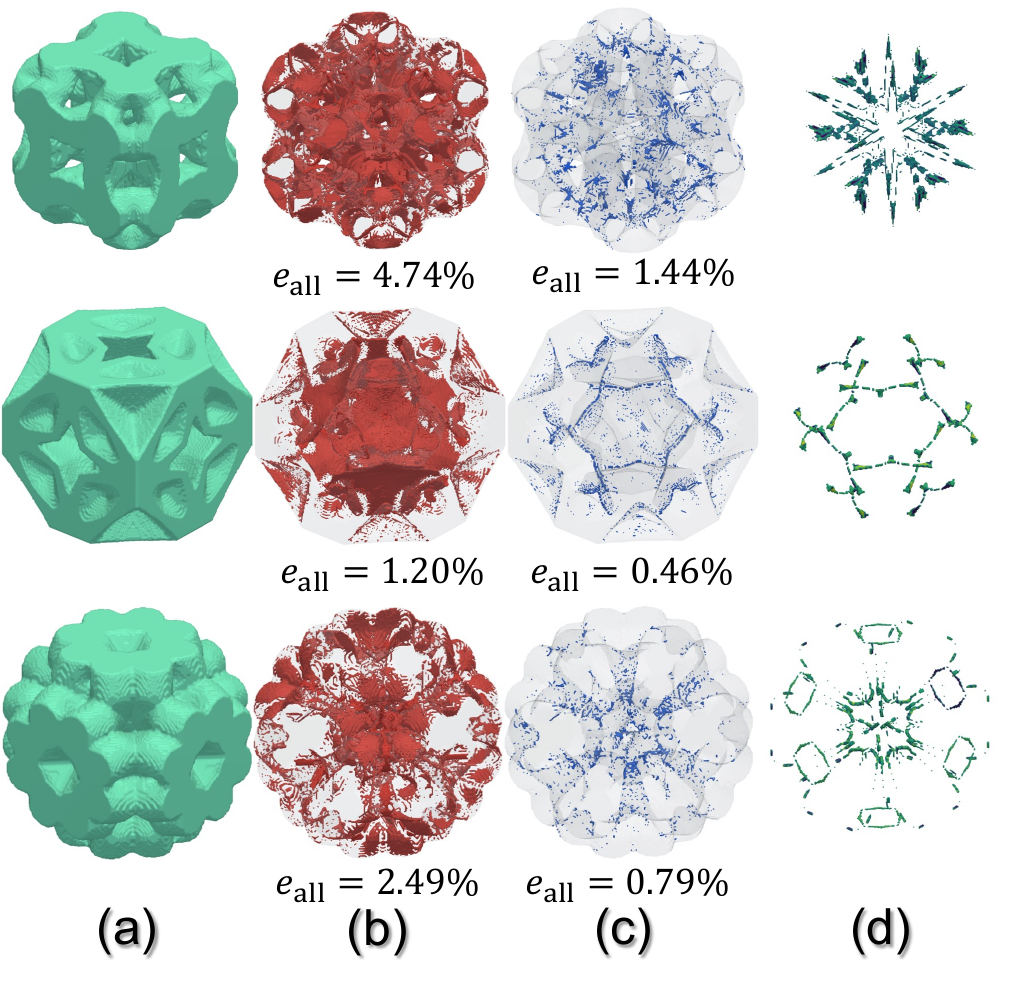}
    \caption{Qualitative ablation of the target-state sensitivity loss on three held-out lattices. Each row shows (a) the paired topology-optimized target, (b) the prediction of the ablated model trained without $\mathcal{L}_{\mathrm{sens}}$, (c) the full SLIM prediction, and (d) regions of high target-state sensitivity computed at the TO target. Values below (b) and (c) report the normalized aggregate TO-target error $e_{\mathrm{all}}^{\mathrm{TO}}$. For all three examples, incorporating $\mathcal{L}_{\mathrm{sens}}$ produces a closer match to the TO target, while the sensitivity maps highlight regions in which residual geometry errors are expected to have larger first-order effects on stiffness.}
\label{fig:sensitivity_ablation}
\end{figure}

To isolate the contribution of the target-state sensitivity loss, we compare the full SLIM model with an otherwise identical model trained without $\mathcal{L}_{\mathrm{sens}}$. As shown in Table~\ref{tab:lightweighting}, adding $\mathcal{L}_{\mathrm{sens}}$ increases the mean relative volume-fraction reduction from $14.54\%$ to $15.90\%$ and the success rate from $78.37\%$ to $90.51\%$, while reducing $e_{\mathrm{all}}^{\mathrm{TO}}$ from $0.831\%$ to $0.636\%$. The component errors in $C_{11}$, $C_{12}$, and $C_{44}$ decrease by $31.0\%$, $29.7\%$, and $6.0\%$, respectively.

Figure~\ref{fig:sensitivity_ablation} provides a qualitative explanation for these improvements. Across the three examples, the full model reduces $e_{\mathrm{all}}^{\mathrm{TO}}$ from $4.74\%$ to $1.44\%$, from $1.20\%$ to $0.46\%$, and from $2.49\%$ to $0.79\%$. The differences between the ablated and full predictions are most evident near the high-sensitivity regions in Fig.~\ref{fig:sensitivity_ablation}(d). This pattern supports the intended role of $\mathcal{L}_{\mathrm{sens}}$: geometric supervision alone does not account for the mechanical consequences of residual errors, whereas the sensitivity loss penalizes their estimated first-order effects on target stiffness.

%% file: tables/lightweighting.tex
\begin{table*}[ht]
\centering
\label{tab:lightweighting}
\small
\setlength{\tabcolsep}{6pt}
\resizebox{\textwidth}{!}{%
\begin{tabular}{lc|ccc|cccc}
\toprule
& \multicolumn{1}{c|}{Success rate}
& \multicolumn{3}{c|}{Volume fraction}
& \multicolumn{4}{c}{Property error} \\
\cmidrule(lr){2-2}
\cmidrule(lr){3-5}
\cmidrule(lr){6-9}
Method
& $r_{\mathrm{succ}}\uparrow$  
& $\mathrm{VF}_{\min}$
& $\mathrm{VF}_{\max}$
& $r_{\mathrm{VF}}\uparrow$
& $e_{\mathrm{all}}^{\mathrm{TO}}\downarrow$
& $e_{11}^{\mathrm{TO}}\downarrow$
& $e_{12}^{\mathrm{TO}}\downarrow$
& $e_{44}^{\mathrm{TO}}\downarrow$ \\
\midrule
TO
& 100.00\%
& 4.25\%
& 60.53\%
& 14.90\%
& n.a.
& n.a.
& n.a.
& n.a. \\

SLIM w/o $\mathcal{L}_{\mathrm{sens}}$
& 78.37\%
& 4.43\%
& 60.51\%
& 14.54\%
& 0.831\%
& 0.741\%
& 1.055\%
& 0.697\% \\

\textbf{SLIM}
& \textbf{90.51\%}
& \textbf{4.20\%}
& \textbf{60.15\%}
& \textbf{15.90\%}
& \textbf{0.636\%}
& \textbf{0.511\%}
& \textbf{0.742\%}
& \textbf{0.655\%} \\

\midrule
SLIM vs. w/o $\mathcal{L}_{\mathrm{sens}}$
& $\mathbf{+15.5\%}$
& $-5.2\%$
& $-0.6\%$
& $+9.4\%$
& $\mathbf{23.4\%}$
& $\mathbf{31.0\%}$
& $\mathbf{29.7\%}$
& $\mathbf{6.0\%}$ \\
\bottomrule
\end{tabular}%
}
\caption{\textbf{Test-set performance and ablation of the
target-state sensitivity loss on test dataset.}
Direct TO provides the paired teacher targets, while
SLIM w/o $\mathcal{L}_{\mathrm{sens}}$ is an otherwise identical
model trained without the target-state sensitivity loss.
$\mathrm{VF}_{\min}$ and $\mathrm{VF}_{\max}$ give the observed
range of output volume fractions, and $r_{\mathrm{VF}}$ is the
mean relative volume-fraction reduction from the input.
The property columns report normalized errors relative to the
paired TO target. A prediction is counted as successful when
$e_{\mathrm{all}}^{\mathrm{TO}}<1.25\%$ and
$\mathrm{VF}^{\mathrm{pred}}<\mathrm{VF}^{\mathrm{in}}$.
For Direct TO, the TO-target errors are not applicable because it
is the reference itself, and its success rate is 100\% by
construction. The final row reports the relative changes obtained
by adding $\mathcal{L}_{\mathrm{sens}}$ to the ablated model;
positive values in the error columns denote relative error
reductions.}
\end{table*}

%% file: figures/testset.tex
\begin{figure}[ht]
    \centering
    \includegraphics[width=\linewidth]
    {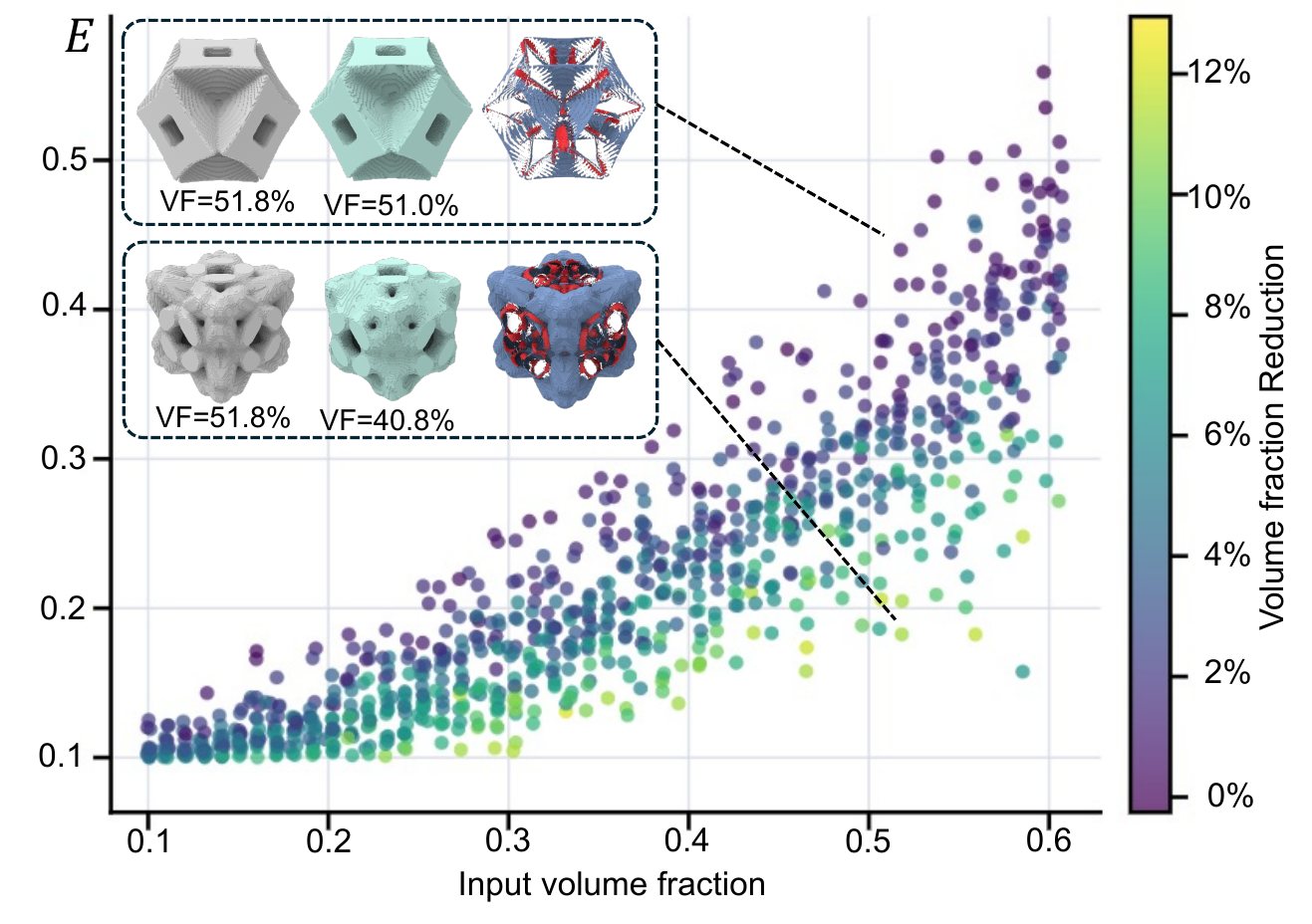}
    \caption{\textbf{Predicted material reduction across the held-out
test distribution.}
Each point represents one test lattice, positioned according to
its input volume fraction $\mathrm{VF}_{\mathrm{in}}$ and effective
Young's modulus $E$. Color encodes the volume-fraction decrease
$\Delta\mathrm{VF}
=\mathrm{VF}_{\mathrm{in}}-\mathrm{VF}_{\mathrm{pred}}$,
expressed in percentage points.
The insets show two lattices with the same input volume fraction
$\mathrm{VF}_{\mathrm{in}}=51.8\%$ but different predicted
reductions. Within each inset, the three renderings show, from left
to right, the input lattice, the SLIM prediction, and their
voxel-wise difference; blue and red denote removed and added
material, respectively. }
    \label{fig:reduction_map}
\end{figure}

%% file: figures/gallary.tex
\begin{figure}[t]
    \centering
    \includegraphics[width=\linewidth]
    {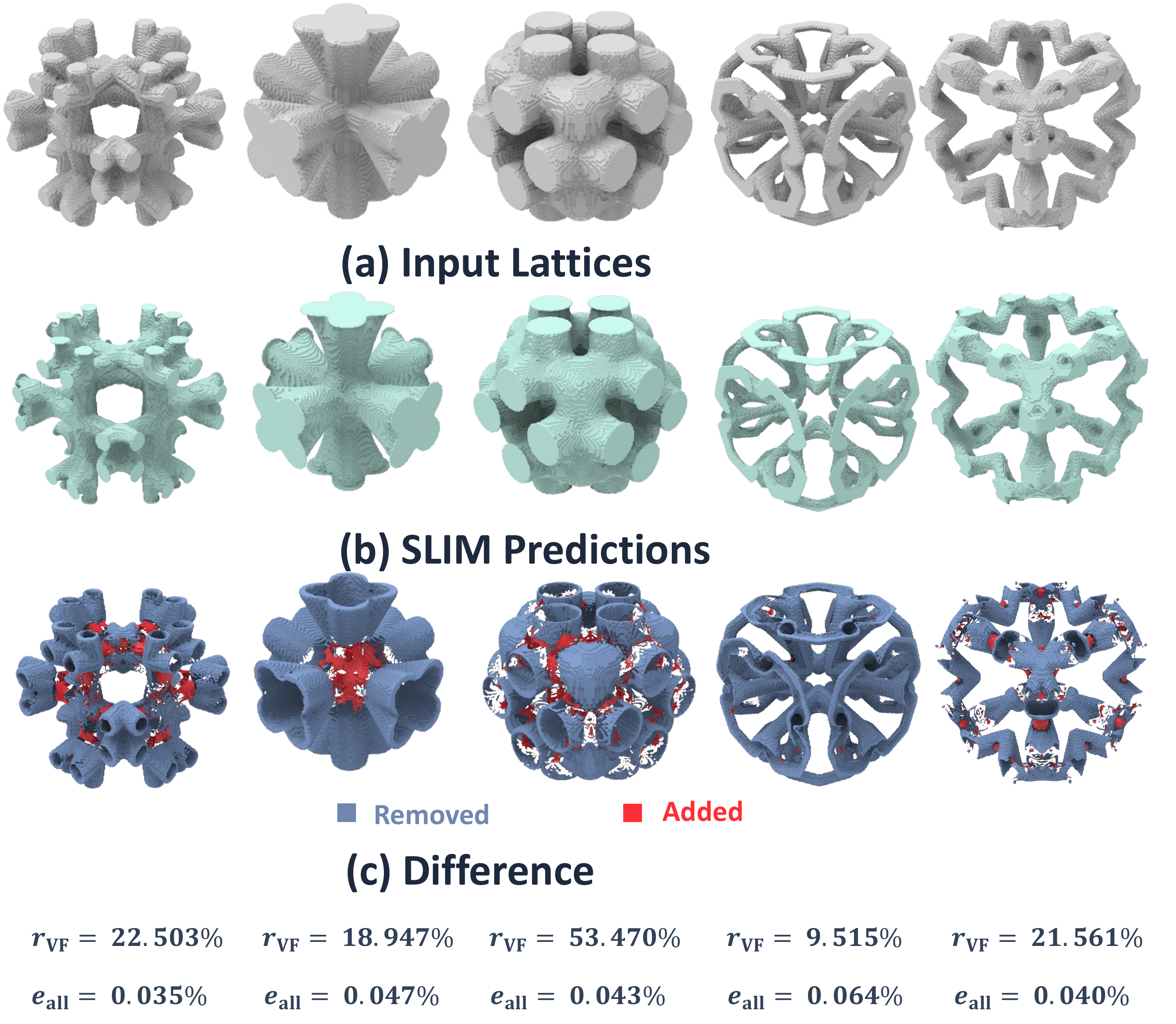}
    \caption{\textbf{Representative single-lattice refinement results.}
Each column shows one example from the held-out test set.
From top to bottom, the rows display the input lattice, the
corresponding SLIM prediction, and the voxel-wise difference
between the prediction and input.
Blue and red denote material removed from and added to the input,
respectively. The values below each column report the relative
volume-fraction reduction $r_{\mathrm{VF}}$ and the normalized
aggregate TO-target error $e_{\mathrm{all}}^{\mathrm{TO}}$.}
    \label{fig:lattice_gallery}
\end{figure}

%% file: figures/assembly.tex
\begin{figure}[ht]
    \centering
    \includegraphics[width=\linewidth]
    {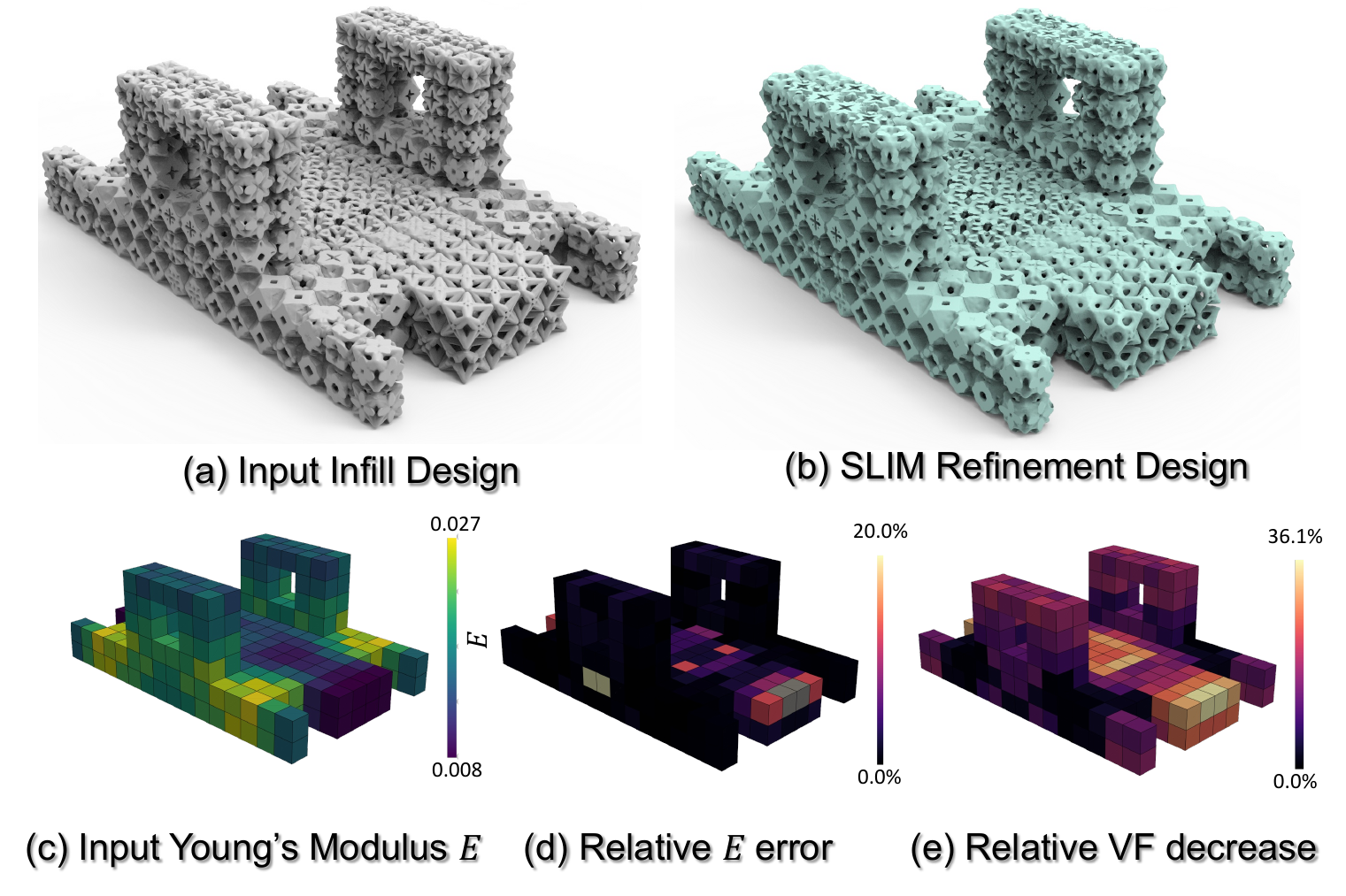}
    \caption{\textbf{Assembly lightweighting.}
    The refined array uses $13.4\%$ less material than the input,
    reducing the volume fraction from $53.5\%$ to $46.3\%$ while
    preserving the overall lattice arrangement.}
    \label{fig:assembly_results}
\end{figure}

%% file: tables/time.tex
\begin{table}[t]
\centering
\label{tab:runtime}
\small
\setlength{\tabcolsep}{6pt}
\begin{tabular}{lcccc}
\toprule
Input VF (\%)& TO iter. & TO (s) & SLIM (s) & Speedup $\uparrow$ \\
\midrule
$[10, 20)$ & $37{,}069.7$ & $2330.7$ & $0.5$ & $4661\times$ \\
$[20, 30)$ & $29{,}869.5$ & $2282.8$ & $0.5$ & $4566\times$ \\
$[30, 40)$ & $28{,}359.2$ & $2453.0$ & $0.5$ & $4906\times$ \\
$[40, 50)$ & $28{,}545.3$ & $2676.1$ & $0.5$ & $5352\times$ \\
$[50, 60]$ & $28{,}996.8$ & $2802.3$ & $0.5$ & $5605\times$ \\
\midrule
\textbf{All} &
$\mathbf{30{,}459.2}$ &
$\mathbf{2516.0}$ &
$\mathbf{0.5}$ &
$\mathbf{5032\times}$ \\
\bottomrule
\end{tabular}
\caption{\textbf{Per-lattice runtime comparison grouped by input
volume-fraction interval.}
All entries are averages measured on identical hardware, with
input volume-fraction intervals expressed in percent.
For topology optimization, ``TO iter.'' is the cumulative number of inner
optimization iterations across all outer volume-fraction trials,
and the reported runtime covers the complete target-generation
procedure. The SLIM runtime includes point sampling, octree
construction, network inference, SDF querying, and binary voxel
reconstruction, but excludes the post-hoc numerical homogenization
used only for evaluation. Speedup is the ratio of the Direct-TO
runtime to the SLIM runtime.}
\end{table}

%% file: src/6-discussion.tex
\section{Conclusion}

We introduced SLIM, an input-conditioned neural operator for
material-efficient refinement of non-parametric 3D lattices. SLIM learns
the geometry transformation represented by boundary- and
property-constrained TO targets and predicts a refined lattice in a single
forward pass. On 906 held-out library lattices, SLIM achieves a mean
relative volume-fraction reduction of $15.90\%$, a lightweighting success
rate of $90.51\%$, and a mean normalized TO-target error
$e_{\mathrm{all}}^{\mathrm{TO}}$ of $0.636\%$. The target-state
sensitivity loss improves the success rate from $78.37\%$ to $90.51\%$
and reduces $e_{\mathrm{all}}^{\mathrm{TO}}$ by $23.4\%$, supporting its
role in stiffness-aware geometric learning. SLIM preserves the prescribed
boundary band by construction and reduces the average refinement-generation
time from $2516.0$~s for direct TO to approximately $0.5$~s, corresponding
to a $5032\times$ speedup. These results establish SLIM as a fast surrogate
for repeated TO-based refinement of in-distribution lattices, rather than
as a constraint-certified replacement for numerical optimization.

\paragraph{Limitations.}
SLIM does not formally guarantee material reduction, connectivity, or
preservation of the input elasticity tensor. Although
$\mathcal{L}_{\mathrm{sens}}$ incorporates first-order mechanical
information, it remains a local training regularizer and does not explicitly
enforce the elasticity constraint. Predictions must therefore be verified
by connectivity checks and numerical homogenization when feasibility is
required; this verification cost is not included in the reported
$0.5$-second generation time. SLIM also inherits the local optimality and
selection bias of its numerical teacher, since the current dataset retains
only inputs for which TO finds a feasible lower-volume target. Finally,
thin members remain vulnerable to SDF and voxelization errors, and the
current evidence is limited to the evaluated lattice distribution, a
$128^3$ representation, and linear homogenized elasticity.

\paragraph{Future Work.}
A natural next step is a hybrid learned-optimization pipeline in which
uncertainty estimates or post-prediction feasibility checks trigger
corrective optimization only when necessary. 
Connectivity- and manufacturability-aware objectives, including explicit
length-scale and additive-manufacturing constraints, may improve the
preservation and printability of thin structural features
~\cite{liu2016manufacturing,gaynor2016overhang,
langelaar2017amfilter}.
Integrating SLIM into multiscale optimization will further require
macroscale feasibility feedback and evaluation under stress, buckling,
nonlinear response, and fabrication constraints.

%% file: src/appendix.tex
\appendix
\section{Homogenization and Property Evaluation}
\label{sec:app_homogenization}

This section gives the homogenization details omitted from the main text.
We follow the standard periodic-cell finite-element formulation for voxelized
microstructures~\cite{andreassen_how_2014}.

\subsection{Periodic Representative Volume Element}
\label{sec:app_periodic_rve}

Let
\(
\Omega=[0,L_x]\times[0,L_y]\times[0,L_z]\subset\mathbb{R}^3
\)
be a periodic representative volume element (RVE), with volume
\(|\Omega|=L_xL_yL_z\) and period vectors
\(\boldsymbol{Y}_{\alpha}=L_{\alpha}\boldsymbol{e}_{\alpha}\),
\(\alpha\in\{x,y,z\}\). In our experiments, each RVE is represented by a
binary field
\(
V\in\{0,1\}^{N\times N\times N}
\)
with \(N=128\), and each voxel is treated as an eight-node hexahedral finite
element. For element \(e\), \(V_e=1\) denotes solid material and \(V_e=0\)
denotes void.

The solid phase is modeled as an isotropic, linearly elastic material with
Young's modulus \(E_0\), Poisson's ratio \(\nu_0\), and elasticity tensor
\(\mathbb{C}_0\). In index notation,
\begin{equation}
  (\mathbb{C}_0)_{ijkl}
  =\lambda_0\delta_{ij}\delta_{kl}
  +\mu_0\left(\delta_{ik}\delta_{jl}
  +\delta_{il}\delta_{jk}\right),
  \label{eq:base_elasticity}
\end{equation}
where
\begin{equation}
  \lambda_0=\frac{E_0\nu_0}{(1+\nu_0)(1-2\nu_0)},
  \qquad
  \mu_0=\frac{E_0}{2(1+\nu_0)}.
  \label{eq:lame_parameters}
\end{equation}

An ersatz material avoids singular systems in void regions. Using the usual
SIMP interpolation, the element tensor is~\cite{sigmund_materials_1994}
\begin{equation}
  \mathbb{C}_e(\rho_e)
  =\left[\rho_{\min}+(1-\rho_{\min})\rho_e^p\right]\mathbb{C}_0,
  \label{eq:simp_interpolation}
\end{equation}
where \(\rho_e\in[0,1]\), \(\rho_{\min}=10^{-5}\), and \(p=3\) for the
continuous topology-optimization variables. When evaluating a binary lattice,
we set \(\rho_e=V_e\); hence, the exponent does not affect the final binary
property evaluation.

\subsection{Periodic Linear-Elasticity Cell Problems}
\label{sec:app_elasticity_cell_problem}

We use the Voigt ordering
\(
I\leftrightarrow(11,22,33,23,13,12)
\)
and the engineering-strain vector
\begin{equation}
  \boldsymbol{\varepsilon}^{\mathrm{V}}
  =\left(
  \varepsilon_{11},\varepsilon_{22},\varepsilon_{33},
  2\varepsilon_{23},2\varepsilon_{13},2\varepsilon_{12}
  \right)^{\mathsf T}.
  \label{eq:voigt_convention}
\end{equation}
The homogenized stiffness tensor is obtained from six independent cell
problems. For loading mode \(I\in\{1,\ldots,6\}\), let
\(\overline{\boldsymbol{\varepsilon}}^{(I)}\) be the unit macroscopic strain
whose Voigt representation is the \(I\)-th canonical basis vector. The total
microscopic displacement is decomposed into an affine part and a periodic
corrector,
\begin{equation}
  \boldsymbol{u}^{(I)}(\boldsymbol{x})
  =\overline{\boldsymbol{\varepsilon}}^{(I)}\boldsymbol{x}
  +\boldsymbol{\chi}^{(I)}(\boldsymbol{x}).
  \label{eq:displacement_decomposition}
\end{equation}
The corrector satisfies the equilibrium equation
\begin{equation}
  \nabla\!\cdot\!\left[
  \mathbb{C}(\boldsymbol{x}):
  \left(
  \overline{\boldsymbol{\varepsilon}}^{(I)}
  +\boldsymbol{\varepsilon}(\boldsymbol{\chi}^{(I)})
  \right)
  \right]=\boldsymbol{0}
  \quad\text{in }\Omega,
  \label{eq:elasticity_cell_problem}
\end{equation}
where
\(
\boldsymbol{\varepsilon}(\boldsymbol{w})
=\tfrac{1}{2}(\nabla\boldsymbol{w}+\nabla\boldsymbol{w}^{\mathsf T})
\)
is the infinitesimal-strain operator. Opposite RVE faces are coupled through
periodic boundary conditions,
\begin{equation}
  \boldsymbol{\chi}^{(I)}
  (\boldsymbol{x}+\boldsymbol{Y}_{\alpha})
  =\boldsymbol{\chi}^{(I)}(\boldsymbol{x}),
  \qquad \alpha\in\{x,y,z\},
  \label{eq:periodic_boundary_condition}
\end{equation}
and uniqueness is enforced by the zero-mean gauge
\begin{equation}
  \left\langle\boldsymbol{\chi}^{(I)}\right\rangle_{\Omega}
  =\frac{1}{|\Omega|}\int_{\Omega}
  \boldsymbol{\chi}^{(I)}\,\mathrm{d}\Omega
  =\boldsymbol{0}.
  \label{eq:zero_mean_gauge}
\end{equation}

For every periodic test function \(\boldsymbol{v}\), the corresponding weak
form is
\begin{equation}
  \int_{\Omega}
  \boldsymbol{\varepsilon}(\boldsymbol{v}):
  \mathbb{C}:
  \boldsymbol{\varepsilon}(\boldsymbol{\chi}^{(I)})
  \,\mathrm{d}\Omega
  =-
  \int_{\Omega}
  \boldsymbol{\varepsilon}(\boldsymbol{v}):
  \mathbb{C}:
  \overline{\boldsymbol{\varepsilon}}^{(I)}
  \,\mathrm{d}\Omega.
  \label{eq:elasticity_weak_form}
\end{equation}

\subsection{Finite-Element Discretization and Effective Tensor}
\label{sec:app_fem_homogenization}

Let \(\mathbf{N}_e\) and \(\mathbf{B}_e\) denote the trilinear shape-function
matrix and strain--displacement matrix of element \(e\), respectively. The
element stiffness matrix and the equivalent load for strain mode \(I\) are
\begin{equation}
  \mathbf{K}_e
  =\int_{\Omega_e}\mathbf{B}_e^{\mathsf T}
  \mathbf{C}_e\mathbf{B}_e\,\mathrm{d}\Omega,
  \qquad
  \mathbf{f}_e^{(I)}
  =-\int_{\Omega_e}\mathbf{B}_e^{\mathsf T}
  \mathbf{C}_e
  \overline{\boldsymbol{\varepsilon}}^{(I)}
  \,\mathrm{d}\Omega,
  \label{eq:element_system}
\end{equation}
where \(\mathbf{C}_e\) is the \(6\times6\) Voigt representation of
\(\mathbb{C}_e\). After assembly and application of the periodic constraints,
the six corrector fields are obtained from
\begin{equation}
  \mathbf{K}\mathbf{d}^{(I)}=\mathbf{f}^{(I)},
  \qquad I=1,\ldots,6.
  \label{eq:global_cell_system}
\end{equation}
All systems are solved using the same stopping criterion
\begin{equation}
  \frac{\|\mathbf{K}\mathbf{d}^{(I)}-\mathbf{f}^{(I)}\|_2}
       {\|\mathbf{f}^{(I)}\|_2}
  \leq \tau_{\mathrm{hom}},
  \label{eq:homogenization_tolerance}
\end{equation}
where \(\tau_{\mathrm{hom}}\) is the prescribed homogenization-solver
tolerance.

The effective tensor follows from Hill--Mandel energy equivalence:
\begin{equation}
  C^{H}_{IJ}
  =\frac{1}{|\Omega|}
  \int_{\Omega}
  \left(
  \overline{\boldsymbol{\varepsilon}}^{(I)}
  +\boldsymbol{\varepsilon}(\boldsymbol{\chi}^{(I)})
  \right):\mathbb{C}:
  \left(
  \overline{\boldsymbol{\varepsilon}}^{(J)}
  +\boldsymbol{\varepsilon}(\boldsymbol{\chi}^{(J)})
  \right)
  \,\mathrm{d}\Omega.
  \label{eq:effective_tensor_continuous}
\end{equation}
Equivalently, its discrete form is
\begin{equation}
  C^{H}_{IJ}
  =\frac{1}{|\Omega|}
  \sum_{e}
  \left(\mathbf{d}^{0(I)}_e+\mathbf{d}^{(I)}_e\right)^{\mathsf T}
  \mathbf{K}_e
  \left(\mathbf{d}^{0(J)}_e+\mathbf{d}^{(J)}_e\right),
  \label{eq:effective_tensor_discrete}
\end{equation}
where \(\mathbf{d}^{0(I)}_e\) contains the element nodal displacements of the
affine unit-strain field and \(\mathbf{d}^{(I)}_e\) contains the periodic
corrector degrees of freedom. This produces
\(\mathbf{C}^{H}\in\mathbb{R}^{6\times6}\) in the Voigt convention of
Eq.~\eqref{eq:voigt_convention}.

\subsection{Cubic Symmetry and Reduced Stiffness Representation}
\label{sec:app_cubic_symmetry}

The homogenization procedure above returns the complete effective stiffness
matrix. For a lattice with cubic symmetry whose symmetry axes are aligned
with the RVE axes, the matrix has only three independent elastic constants in
the present Voigt convention~\cite{xue_mind_2025}:
\begin{equation}
  \mathbf{C}^{H}_{\mathrm{cub}}
  =
  \begin{pmatrix}
    C^{H}_{11} & C^{H}_{12} & C^{H}_{12} & 0 & 0 & 0 \\
    C^{H}_{12} & C^{H}_{11} & C^{H}_{12} & 0 & 0 & 0 \\
    C^{H}_{12} & C^{H}_{12} & C^{H}_{11} & 0 & 0 & 0 \\
    0 & 0 & 0 & C^{H}_{44} & 0 & 0 \\
    0 & 0 & 0 & 0 & C^{H}_{44} & 0 \\
    0 & 0 & 0 & 0 & 0 & C^{H}_{44}
  \end{pmatrix}.
  \label{eq:cubic_stiffness_matrix}
\end{equation}
Here \(C^{H}_{11}\) denotes the three equal normal-stiffness entries,
\(C^{H}_{12}\) the three equal normal coupling entries, and \(C^{H}_{44}\)
the three equal shear entries. The sensitivity loss and component-wise
evaluation use these three independent components, so the vector introduced
in the main paper has \(K=3\) and is ordered as
\begin{equation}
  \mathbf{c}^{H}
  =
  \left(C^{H}_{11},C^{H}_{12},C^{H}_{44}\right)^{\mathsf T}.
  \label{eq:cubic_component_vector}
\end{equation}
For compactness, the component-wise metrics in the main paper suppress the
superscript \(H\), so, for example, \(C_{11}^{\mathrm{pred}}\) denotes
\(C_{11}^{H}(V^{\mathrm{pred}})\).
This is a symmetry-based reduction of the representation, not a reduction of
the cell problems: the six load cases are still solved to obtain the
homogenized response before the selected components are collected. The
topology-optimization objective and feasibility test continue to compare the
complete \(6\times6\) homogenized tensors.

For exact cubic symmetry, the corresponding compliance matrix is also cubic.
Consequently, the directional averages in Eq.~\eqref{eq:average_properties}
reduce to the usual cubic expressions
\begin{equation}
  \begin{aligned}
    E
    &=\frac{(C^{H}_{11}-C^{H}_{12})(C^{H}_{11}+2C^{H}_{12})}
      {C^{H}_{11}+C^{H}_{12}}, \\
    \nu&=\frac{C^{H}_{12}}{C^{H}_{11}+C^{H}_{12}},
    \qquad G=C^{H}_{44}.
  \end{aligned}
  \label{eq:cubic_derived_properties}
\end{equation}
For structures that do not satisfy cubic symmetry, or for numerical results
with appreciable symmetry-breaking entries, we retain the full
\(6\times6\) tensor in the homogenization and target-generation calculations
rather than replacing it by the three-constant form.

\subsection{Derived Properties and Evaluation Metrics}
\label{sec:app_property_metrics}

Let
\begin{equation}
  \mathbf{S}^{H}=(\mathbf{C}^{H})^{-1}
  \label{eq:effective_compliance}
\end{equation}
be the effective compliance matrix. Under uniaxial loading along principal
direction \(i\in\{1,2,3\}\), the directional Young's modulus and Poisson's
ratio are
\begin{equation}
  E_i=\frac{1}{S^{H}_{ii}},
  \qquad
  \nu_{ij}=-\frac{S^{H}_{ji}}{S^{H}_{ii}},
  \quad i,j\in\{1,2,3\},\;i\neq j,
  \label{eq:directional_properties}
\end{equation}
where \(\nu_{ij}\) denotes transverse strain in direction \(j\) under uniaxial
stress in direction \(i\). The scalar values reported in the main paper are
directional averages,
\begin{equation}
  E=\frac{1}{3}\sum_{i=1}^{3}E_i,
  \qquad
  \nu=\frac{1}{6}\sum_{i=1}^{3}
  \sum_{\substack{j=1\\j\neq i}}^{3}\nu_{ij}.
  \label{eq:average_properties}
\end{equation}
Because the evaluated lattices are not assumed to be isotropic, these two
scalars are only interpretable property summaries.

For a binary lattice \(V\), its volume fraction is
\begin{equation}
  \operatorname{VF}(V)=\frac{1}{N^3}\sum_{e=1}^{N^3}V_e.
  \label{eq:volume_fraction}
\end{equation}
Given an input lattice \(V^{\mathrm{in}}\) and its final SLIM output
\(V^{\mathrm{pred}}\), the relative volume reduction is
\begin{equation}
  r_{\mathrm{VF}}
  =\frac{\operatorname{VF}(V^{\mathrm{in}})
  -\operatorname{VF}(V^{\mathrm{pred}})}
  {\operatorname{VF}(V^{\mathrm{in}})}.
  \label{eq:relative_volume_reduction}
\end{equation}
Stiffness fidelity is reported only through the TO-target metrics defined in Eqs. (16) and (17) in main text:
\(e_{11}^{\mathrm{TO}}\), \(e_{12}^{\mathrm{TO}}\),
\(e_{44}^{\mathrm{TO}}\), and \(e_{\mathrm{all}}^{\mathrm{TO}}\).

All quantities are recomputed from the final binary geometry after SDF
thresholding and deterministic replacement of the prescribed boundary band.
Elastic-property preservation remains a post-hoc numerical measurement rather
than a formal guarantee of the learned operator.

\section{Topology-Optimization Target Generation}
\label{sec:app_to_target_generation}


The training targets are generated by adapting the GPU-based inverse-
homogenization framework of~\cite{zhang_optimized_2023}, building on the
classical homogenization-based topology-optimization formulation~\cite{bendsoe_kikuchi_1988}
and sensitivity-analysis perspective~\cite{allaire_structural_2004}. We keep the full
optimization details here because the main text only summarizes the pipeline.

\subsection{Target Tensor and Ideal Refinement Problem}
\label{sec:app_to_ideal_problem}

For an input binary lattice
\(
V^{\mathrm{in}}\in\{0,1\}^{N\times N\times N}
\), we first compute its homogenized elasticity tensor using the procedure in
Section~\ref{sec:app_homogenization}:
\begin{equation}
  \mathbf{C}^{\mathrm{tar}}
  :=\mathbf{C}^{H}\!\left(V^{\mathrm{in}}\right).
  \label{eq:to_target_tensor}
\end{equation}
This target remains fixed throughout target generation; it is not updated from
intermediate optimized structures. Let \(\mathcal{B}\) denote the prescribed
boundary-interface band whose voxels must remain identical to those of the
input. The ideal binary refinement problem is
\begin{equation}
  \begin{aligned}
  \min_{V\in\{0,1\}^{N\times N\times N}}\quad
    & \operatorname{VF}(V) \\
  \text{s.t.}\quad
    & \frac{\left\|\mathbf{C}^{H}(V)-\mathbf{C}^{\mathrm{tar}}\right\|_F}
    {\left\|\mathbf{C}^{\mathrm{tar}}\right\|_F}
    \leq \epsilon_C, \\
    & V\vert_{\mathcal{B}}
    =V^{\mathrm{in}}\vert_{\mathcal{B}},
  \end{aligned}
  \label{eq:to_ideal_binary_problem}
\end{equation}
where \(\epsilon_C\) is the prescribed relative elasticity-tensor tolerance.
Because Eq.~\eqref{eq:to_ideal_binary_problem} is a large, non-convex discrete
optimization problem, we approximate it by combining a continuous inner solve
with an outer search over the admissible volume fraction.


\subsection{Inner Tensor-Matching Optimization}
\label{sec:app_to_inner_problem}

For a candidate volume-fraction bound \(v\), the binary field is relaxed to a
continuous density field
\(
\boldsymbol{\rho}=\{\rho_e\}_{e=1}^{M}
\), where \(M=N^3\) and \(\rho_e\in[0,1]\). A periodic density filter is
applied before material interpolation, following the implementation of~\cite{zhang_optimized_2023}. To simplify notation, \(\rho_e\) below
denotes the resulting physical density. The element elasticity tensor follows
the SIMP model in Eq.~\eqref{eq:simp_interpolation}.

We define the normalized tensor-matching objective
\begin{equation}
  J_C(\boldsymbol{\rho})
  =\frac{1}{2}
  \frac{\left\|
  \mathbf{C}^{H}(\boldsymbol{\rho})-\mathbf{C}^{\mathrm{tar}}
  \right\|_F^2}
  {\left\|\mathbf{C}^{\mathrm{tar}}\right\|_F^2}.
  \label{eq:to_tensor_matching_objective}
\end{equation}
The inner optimization problem for the trial bound \(v\) is
\begin{equation}
  \begin{aligned}
  \boldsymbol{\rho}^{\star}(v)
  \in\arg\min_{\boldsymbol{\rho}}\quad
    & J_C(\boldsymbol{\rho}) \\
  \text{s.t.}\quad
    & \frac{1}{M}\sum_{e=1}^{M}\rho_e\leq v, \\
    & 0\leq\rho_e\leq1,\qquad e=1,\ldots,M, \\
    & \rho_e=V^{\mathrm{in}}_e,
      \qquad e\in\mathcal{B}.
  \end{aligned}
  \label{eq:to_inner_problem}
\end{equation}
The boundary densities are overwritten after filtering and are excluded from
the design-variable update, so the interface constraint is always satisfied.

At each inner iteration, the six periodic elasticity systems are solved and
\(\mathbf{C}^{H}(\boldsymbol{\rho})\) is evaluated. The derivative of the
matching objective is
\begin{equation}
   \frac{\partial J_C}{\partial\rho_e}
  =\frac{1}{\left\|\mathbf{C}^{\mathrm{tar}}\right\|_F^2}
  \sum_{I=1}^{6}\sum_{J=1}^{6}
  \left(C^{H}_{IJ}-C^{\mathrm{tar}}_{IJ}\right)
  \frac{\partial C^{H}_{IJ}}{\partial\rho_e}.
  \label{eq:to_objective_sensitivity}
\end{equation}
Under the SIMP interpolation of Eq.~\eqref{eq:simp_interpolation}, the
elementwise tensor sensitivity can be written as
\begin{equation}
  \frac{\partial C^{H}_{IJ}}{\partial\rho_e}
  =\frac{(1-\rho_{\min})p\rho_e^{p-1}}{|\Omega|}
  \left(\mathbf{q}^{(I)}_e\right)^{\mathsf T}
  \mathbf{K}_{0,e}
  \mathbf{q}^{(J)}_e,
  \label{eq:to_tensor_sensitivity}
\end{equation}
where \(\mathbf{K}_{0,e}\) is the stiffness matrix of a fully solid element
and
\(
\mathbf{q}^{(I)}_e
=\mathbf{d}^{0(I)}_e+\mathbf{d}^{(I)}_e
\)
is the element displacement associated with the total microscopic strain of
mode \(I\), using the sign convention of
Eq.~\eqref{eq:effective_tensor_discrete}. Gradients are propagated through the
periodic density filter. We use the method of moving asymptotes (MMA) to update
the free densities subject to the trial volume bound in
Eq.~\eqref{eq:to_inner_problem}.

The same analytic derivative is used to construct the target-state
Jacobian for network training. Specifically,
Eq.~\eqref{eq:to_tensor_sensitivity} is evaluated once at the finalized
TO target \(\boldsymbol{\rho}^{\mathrm{TO}}\), after restoring the
prescribed boundary band. The rows corresponding to
\(C^{H}_{11}\), \(C^{H}_{12}\), and \(C^{H}_{44}\) are collected in the
order given by Eq.~\eqref{eq:cubic_component_vector}. The chain rule
through the periodic density filter is included when forming these
derivatives. The resulting \(K\times M\) Jacobian is exported and held
fixed during training; no finite-difference evaluation, online
homogenization solve, or Jacobian update is required.

\subsection{Binary Feasibility Test}
\label{sec:app_to_binary_check}

The continuous objective value is not used by itself to declare a trial volume
feasible. At selected inner iterations \(k\), the filtered density is
binarized as
\begin{equation}
  V^{(k)}_e(v)
  =\mathbb{I}\!\left[\rho^{(k)}_e(v)\geq\eta_{\mathrm{bin}}\right],
  \label{eq:to_binarization}
\end{equation}
where \(\eta_{\mathrm{bin}}\) is the fixed binarization threshold. The boundary
band is then restored exactly:
\begin{equation}
  V^{(k)}_e(v)\leftarrow V^{\mathrm{in}}_e,
  \qquad e\in\mathcal{B}.
  \label{eq:to_restore_boundary}
\end{equation}
We recompute the homogenized tensor of this binary structure rather than using
the tensor of the continuous density field. A binary checkpoint is feasible at
volume bound \(v\) if
\begin{equation}
  \begin{aligned}
    \operatorname{VF}\!\left(V^{(k)}(v)\right)
    &\leq v+\epsilon_{\mathrm{VF}}, \\
    \frac{\left\|\mathbf{C}^{H}\!\left(V^{(k)}(v)\right)
    -\mathbf{C}^{H}\!\left(V^{\mathrm{in}}\right)\right\|_F}
    {\left\|\mathbf{C}^{H}\!\left(V^{\mathrm{in}}\right)\right\|_F}
    &\leq\epsilon_C.
  \end{aligned}
  \label{eq:to_binary_feasibility}
\end{equation}
where \(\|\cdot\|_F\) is evaluated under the same Voigt convention for both
tensors, and \(\epsilon_{\mathrm{VF}}\) allows only numerical volume-rounding
error. Among
all feasible checkpoints generated for a trial \(v\), we retain the one with
the smallest binary volume fraction.

\subsection{Outer Bisection over Volume Fraction}
\label{sec:app_to_bisection}

The upper bound is initialized as
\begin{equation}
  v_{\mathrm{hi}}=\operatorname{VF}(V^{\mathrm{in}}),
  \label{eq:to_initial_upper_bound}
\end{equation}
which is feasible because the input itself exactly realizes
\(\mathbf{C}^{\mathrm{tar}}\). A geometric lower bound is given by the solid
material fixed inside the boundary band,
\begin{equation}
  v_{\mathrm{lo}}
  =\frac{1}{M}\sum_{e\in\mathcal{B}}V^{\mathrm{in}}_e.
  \label{eq:to_initial_lower_bound}
\end{equation}
For each outer iteration, we set
\(
v_{\mathrm{mid}}=(v_{\mathrm{lo}}+v_{\mathrm{hi}})/2
\)
and solve Eq.~\eqref{eq:to_inner_problem}. If at least one binary checkpoint
satisfies Eq.~\eqref{eq:to_binary_feasibility}, the trial is marked feasible,
the checkpoint is stored, and
\(v_{\mathrm{hi}}\leftarrow v_{\mathrm{mid}}\). Otherwise,
\(v_{\mathrm{lo}}\leftarrow v_{\mathrm{mid}}\). The search terminates when
\begin{equation}
  v_{\mathrm{hi}}-v_{\mathrm{lo}}\leq\epsilon_v
  \label{eq:to_bisection_tolerance}
\end{equation}
or when the maximum number of outer iterations is reached.

\begin{algorithm}[t]
  \caption{Boundary- and property-constrained target generation}
  \label{alg:to_target_generation}
  \begin{algorithmic}[1]
    \Require Input lattice \(V^{\mathrm{in}}\), boundary band \(\mathcal{B}\),
      tensor tolerance \(\epsilon_C\), and bisection tolerance \(\epsilon_v\)
    \State \(\mathbf{C}^{\mathrm{tar}}
      \gets\mathbf{C}^{H}(V^{\mathrm{in}})\)
    \State \(v_{\mathrm{hi}}\gets\operatorname{VF}(V^{\mathrm{in}})\)
    \State \(v_{\mathrm{lo}}\gets M^{-1}
      \sum_{e\in\mathcal{B}}V^{\mathrm{in}}_e\)
    \State \(\mathcal{Q}_{\mathrm{feas}}\gets\varnothing\)
    \While{\(v_{\mathrm{hi}}-v_{\mathrm{lo}}>\epsilon_v\)}
      \State \(v_{\mathrm{mid}}
        \gets(v_{\mathrm{lo}}+v_{\mathrm{hi}})/2\)
      \State Solve Eq.~\eqref{eq:to_inner_problem} at \(v_{\mathrm{mid}}\)
      \State Binarize selected iterates and recompute their tensors
      \If{a checkpoint satisfies Eq.~\eqref{eq:to_binary_feasibility}}
        \State Add the lowest-volume feasible checkpoint to
          \(\mathcal{Q}_{\mathrm{feas}}\)
        \State \(v_{\mathrm{hi}}\gets v_{\mathrm{mid}}\)
      \Else
        \State \(v_{\mathrm{lo}}\gets v_{\mathrm{mid}}\)
      \EndIf
    \EndWhile
    \State \Return the member of \(\mathcal{Q}_{\mathrm{feas}}\) with the
      smallest binary volume fraction
  \end{algorithmic}
\end{algorithm}

Finally, the topology-optimization target is selected as
\begin{equation}
  V^{\mathrm{TO}}
  =\arg\min_{V\in\mathcal{Q}_{\mathrm{feas}}}
  \operatorname{VF}(V).
  \label{eq:to_final_target}
\end{equation}
An input is retained as a paired training sample only if
\(\mathcal{Q}_{\mathrm{feas}}\neq\varnothing\) and the selected target has a
strictly smaller binary volume fraction than the input. The final pair is
\((V^{\mathrm{in}},V^{\mathrm{TO}})\).

After applying these retention criteria, the final dataset contains
9259 input--target lattice pairs. The data are split
by input-lattice identity into 7440 training pairs,
913 validation pairs, and 906 held-out test pairs.

The nested procedure searches for the smallest \emph{observed feasible}
volume fraction. The continuous feasible sets are nested with respect to the
volume bound, which motivates bisection. Nevertheless, the inner problem is
non-convex and is solved only to a local optimum; binarization can also alter
both volume and stiffness. Therefore, \(V^{\mathrm{TO}}\) is a numerical
teacher solution and is not claimed to be a global minimizer of
Eq.~\eqref{eq:to_ideal_binary_problem}.

\section{SLIM Architecture and Implicit Representation}
\label{sec:app_architecture}
SLIM uses the dual-octree GraphOUNet described in the main paper, following
the adaptive volumetric representation of~\cite{wang_dualoctree_2022}. The input
octree has maximum depth 6 and full depth 3. The encoder and decoder contain
three GraphResBlocks at each level, with encoder widths
$(32,64,128,256)$ and decoder widths $(256,128,64,32)$. The network uses
batch normalization, ReLU activations, basic residual blocks, and bottleneck
width 4. The input feature has three channels and the graph contains seven
edge types.

The input lattice is converted into an oriented point set and encoded by an
adaptive octree. The continuous signed-distance field is an implicit shape
representation related to DeepSDF~\cite{park_deepsdf_2019}. Coordinates are normalized using point scale $0.5$. During
each update, 10,000 SDF queries are sampled from the preprocessed cache.
Data augmentation is disabled and training batches are shuffled.

The geometry objective consists of the octree split loss, SDF regression, and
SDF-gradient regression:
\begin{equation}
\mathcal{L}_{\mathrm{geo}}
=
\mathcal{L}_{\mathrm{split}}
+
\lambda_{\mathrm{sdf}}\mathcal{L}_{\mathrm{sdf}}
+
\lambda_{\mathrm{grad}}\mathcal{L}_{\mathrm{grad}}.
\label{eq:app_geometry_loss}
\end{equation}
Thus, the notation of the main paper corresponds to
\(\lambda_{\mathrm{sdf}}=200\) and \(\lambda_{\mathrm{grad}}=1\).
The network predicts a continuous SDF. At inference, the predicted field is
evaluated on a dense $128^3$ grid and thresholded at zero to obtain a binary
lattice.

For both Baseline and SLIM, the outer one-voxel interface shell is copied from
the input occupancy after SDF thresholding. This operation is inference-only
and is not an SLIM-specific advantage. No topology-optimization or LIVE3D solve
is executed during the network forward pass.

\section{Training and Reproducibility Details}
\label{sec:app_training}

The Baseline is trained from random initialization for 1,000
epochs using AdamW with learning rate $10^{-3}$, weight decay $10^{-2}$,
batch size 16, polynomial learning-rate decay to $10^{-4}$, and a 20-epoch
warm-up.

SLIM is initialized from the resulting Baseline checkpoint. It uses the same
network and data representation, AdamW weight decay $10^{-2}$, learning rate
$1.5\times10^{-5}$ with cosine decay, batch size 8, frozen batch-normalization
statistics, and a 300-epoch fine-tuning budget. Checkpoints are saved every
20 epochs. The reported SLIM checkpoint is selected at epoch 200 using only the
fixed 913-sample validation split. The 906-sample test split remains locked
until the final comparison.

During development,
\[
\lambda_J\in
\{3\times10^{-4},6\times10^{-4},10^{-3}\}
\]
was evaluated using the fixed validation split. Candidate ranking used
the label-referenced validation errors recorded by the training pipeline,
including the absolute errors in \(C^{H}_{11}\) and \(C^{H}_{12}\). IoU was
reported as a diagnostic but was not used for model selection. The test split
was not inspected during this search.

The Code and Data Supplement contains the SLIM/OGNN and LIVE3D source code,
three example input--target pairs, the trained \texttt{slim.pth} checkpoint,
and configuration files. The README explains how
to configure the environment and run inference on the included examples.
Full-dataset training and 906-sample evaluation require the complete dataset
and a separately compiled LIVE3D executable.

\paragraph{Matched-budget ablation.}
The direct Baseline--SLIM comparison is not a matched-compute ablation:
Baseline is trained for 1,000 epochs from random initialization, whereas SLIM
is initialized from that checkpoint and fine-tuned for up to 300 additional
epochs. Therefore, the difference between these two rows cannot be attributed
solely to the sensitivity loss $\mathcal{L}_{\mathrm{sens}}$.

To isolate the contribution of $\mathcal{L}_{\mathrm{sens}}$, we use a matched
control initialized from the identical 1,000-epoch Baseline checkpoint. The
control uses the same 300-epoch schedule, learning rate, batch size, random
seed, frozen batch-normalization statistics, data fields, and checkpoint
selection protocol as SLIM. Its only intentional difference is that the
sensitivity-loss weight is set to zero.

On the locked 906-sample test split, evaluated with the same criteria as the
main paper, the matched control obtains a success rate
$r_{\mathrm{succ}}=82.78\%$, mean relative volume reduction
$r_{\mathrm{VF}}=15.54\%$, and mean Yang range-normalized property error
$e_{\mathrm{all}}=0.749\%$. SLIM obtains
$r_{\mathrm{succ}}=90.51\%$, $r_{\mathrm{VF}}=15.90\%$, and
$e_{\mathrm{all}}=0.636\%$, respectively. Here success requires both
$\mathrm{VF}^{\mathrm{out}}<\mathrm{VF}^{\mathrm{in}}$ and
$e_{\mathrm{all}}<1.25\%$, exactly as defined in the main-paper evaluation.

Thus, the original Baseline--SLIM comparison should be interpreted as an
end-to-end training comparison, while the matched control provides the fair
ablation of $\mathcal{L}_{\mathrm{sens}}$.

For the matched-budget ablation, the paired difference is defined as
\[
d_n=m_n^{\mathrm{SLIM}}-m_n^{\mathrm{Control}},
\qquad n=1,\ldots,906.
\]
Because lower error is better, we use a one-sided paired Wilcoxon
signed-rank test with the alternative hypothesis that the SLIM error is
smaller than the matched-control error. Zero differences are handled using
the Pratt convention. No test-specific sample removal or outlier filtering is
applied. All data, paired differences, and statistical calculations use
\texttt{float64} arithmetic. Very small $p$-values such as $10^{-51}$ or
$10^{-54}$ indicate numerical magnitude.

\begin{table}[t]
\centering
\caption{Paired one-sided Wilcoxon signed-rank tests comparing SLIM with the
matched control on the locked 906-sample test set. All metrics are the
training-range-normalized errors used in the main paper and are computed
against the Direct-TO label; smaller values are better.}
\label{tab:app_wilcoxon}
\small
\setlength{\tabcolsep}{6pt}
\begin{tabular}{lc}
\toprule
Metric & $p$-value \\
\midrule
$e_{\mathrm{all}}^{\mathrm{TO}}$ & $3.43\times10^{-19}$ \\
$e_{11}^{\mathrm{TO}}$           & $2.37\times10^{-17}$ \\
$e_{12}^{\mathrm{TO}}$           & $2.57\times10^{-51}$ \\
$e_{44}^{\mathrm{TO}}$           & $9.92\times10^{-1}$ \\
\bottomrule
\end{tabular}
\end{table}

The matched-budget test provides evidence that SLIM reduces the aggregate
normalized error $e_{\mathrm{all}}^{\mathrm{TO}}$ and the two dominant axial
components $e_{11}^{\mathrm{TO}}$ and $e_{12}^{\mathrm{TO}}$ relative to the
control. The result for the shear component $e_{44}^{\mathrm{TO}}$ is not
statistically significant, so no uniform improvement is claimed for every
stiffness component.

\section{Additional Quantitative Results}
\label{sec:app_additional_results}

We use the same notation as in the main paper and report the three
training-range-normalized errors relative to the Direct-TO target,
$e_{11}^{\mathrm{TO}}$, $e_{12}^{\mathrm{TO}}$, and
$e_{44}^{\mathrm{TO}}$, together with their arithmetic mean
$e_{\mathrm{all}}^{\mathrm{TO}}$.

For each metric $m$, the paired difference is defined as
\[
d_n=m_n^{\mathrm{SLIM}}-m_n^{\mathrm{Baseline}},
\qquad n=1,\ldots,906.
\]
Because lower error is better, we use a one-sided paired Wilcoxon
signed-rank test with the alternative hypothesis that the SLIM error is
smaller than the Baseline error. Zero differences are handled using the
Pratt convention. No test-specific sample removal or outlier filtering is
applied.

\begin{table}[t]
\centering
\caption{Paired one-sided Wilcoxon signed-rank tests comparing SLIM with
Baseline on the locked 906-sample test set. All metrics are computed against
the Direct-TO label; smaller values are better.}
\label{tab:app_wilcoxon}
\small
\setlength{\tabcolsep}{6pt}
\begin{tabular}{lc}
\toprule
Metric & $p$-value \\
\midrule
$e_{\mathrm{all}}^{\mathrm{TO}}$ & $3.64\times10^{-24}$ \\
$e_{11}^{\mathrm{TO}}$           & $2.72\times10^{-29}$ \\
$e_{12}^{\mathrm{TO}}$           & $1.35\times10^{-54}$ \\
$e_{44}^{\mathrm{TO}}$           & $3.71\times10^{-1}$ \\
\bottomrule
\end{tabular}
\end{table}

The test provides strong evidence that SLIM reduces the aggregate normalized
error $e_{\mathrm{all}}^{\mathrm{TO}}$ and the two dominant axial components
$e_{11}^{\mathrm{TO}}$ and $e_{12}^{\mathrm{TO}}$. The result for the shear
component $e_{44}^{\mathrm{TO}}$ is not statistically significant, so no
uniform improvement is claimed for every stiffness component.